\documentclass{pasj00}
\usepackage{graphicx}
\usepackage{natbib}

\def\evar{{\cal E}}
\def\evarh{{\cal E}_h}
\def\mbh{M_{bh}}
\def\rc{r_c}
\def\rcusp{r_{cusp}}
\def\rh{r_h}
\def\rhoc{\rho_c}
\def\rhoh{\rho_h}
\def\tr{t_{r}}
\def\trh{t_{rh}}
\def\vc{v_c}
\def\vh{v_h}

\begin{document}
\SetRunningHead{Heggie, Hut, Mineshige, Makino \& Baumgardt}{The Core Radius of a Star Cluster Containing a Black Hole}
\Received{2006/12/01}
\Accepted{2007/04/10}

\title{The Core Radius of a Star Cluster Containing a Massive Black Hole}

\author{Douglas C. \textsc{Heggie}}
\affil{School of Mathematics and Maxwell Institute for Mathematical
  Sciences, University of Edinburgh, \\
King's Buildings, Edinburgh EH9 3JZ, UK}\email{d.c.heggie@ed.ac.uk}
\author{Piet \textsc{Hut}}
\affil{Institute for Advanced Study, Princeton, NJ 08540, USA}\email{piet@ias.edu}
\author{Shin \textsc{Mineshige}}
\affil{Yukawa Institute for Theoretical Physics, Kyoto University, Sakyo-ku,
Kyoto 606-8502, Japan}\email{minesige@yukawa.kyoto-u.ac.jp}
\author{Jun {\sc Makino}}
\affil{Center for Computational Astrophysics,
National Astronomical Observatory of Japan,\\
2-21-1 Ohsawa, Mitaka, Tokyo 181-8588, Japan}
\email{makino@th.nao.ac.jp}
\and
\author{Holger \textsc{Baumgardt}}
\affil{Argelander Institute for Astronomy (AIfA), Auf dem H\"ugel 71, 53121 Bonn, Germany}
\email{holger@astro.uni-bonn.de}


%


\maketitle

\begin{abstract}
We present a theoretical framework which establishes how the core
radius of a star cluster varies with the mass of an assumed central
black hole. 
Our result is that $r_c/r_h\propto (\mbh/M)^{3/4}$ when the system is
well relaxed.  The theory compares favourably 
with a number of simulations of this problem{, which extend to
  black hole masses of order 10\% of the cluster mass.  Though
  strictly limited
  as yet to clusters with stars of equal mass, our conclusion
  strengthens the view that clusters with large core radii are the
  most promising candidates in which to find a massive black hole.}
\end{abstract}


\section{Introduction}

Though the existence of intermediate-mass black holes in star clusters
remains controversial, our theoretical understanding of the problem
has advanced on two fronts.  First, important conclusions have been
reached from purely theoretical considerations, including the density
profile of the cusp surrounding the black hole \citep{BW1976,SL1976}.  This can be understood as the
response of the stellar distribution to the steady transport of stars
into the vicinity of the black hole, where they are tidally disrupted,
contributing to the growth of the black hole.  Two-body relaxation is
the vital process which controls the rate at which this flux can be
sustained.  
Second, a succession of simulations have added much detail
to the general theoretical picture.  These simulations have been based
on a variety of techniques: Monte Carlo methods based on a
Fokker-Planck treatment of relaxation with an anisotropic distribution
of velocities \citep{SM1978,DS1982,FB2002};
finite-difference solution of the Fokker-Planck equation for both
anisotropic \citep{CK1978} and 
isotropic distribution functions \citep{Mu1991}; gas models,
in which relaxation is mimicked by a suitably crafted form for thermal
conductivity in a self-gravitating gas \citep{AS2004}; a tree code \citep{Ar1997}; and, most
recently, direct $N$-body simulations \citep{Ba2004a,Ba2004b,Ba2005,TAMH2007}.

In many of these studies, emphasis is given to the details of the cusp and the growth of the
black hole, and less attention is paid to the evolution of the
star cluster.  In this letter we shall show that rather simple
considerations allow us to predict also the evolution of the structure
of the cluster,  in particular its core radius.  The basic
idea is a familiar one in the stellar dynamics of star clusters, where
it was introduced by \citet{He1975}.  This idea is that the flux of
energy from the centre of a star cluster must reach an equilibrium
with the flux (by relaxation) across the outer parts of the cluster,
conventionally taken to be the half-mass radius.  If too much energy
is generated the core of the star cluster must expand to quench the
generation of energy there, no matter what is the mechanism of energy
generation; while insufficient generation of energy leads to the
familiar process of core collapse.  For the case of a central black
hole, energy is generated as stars fall by relaxation towards the
radius around the black hole at which they are disrupted\footnote{Our arguments also apply, in principle, to a purely classical,
  point-mass idealisation, in which stars simply accumulate around
  the black hole at small distances.}.  Though the
link between the energy generated in the cusp and the expansion of the
system has been discussed previously (see Sec.\ref{sec:discussion}),
we believe that the theoretical estimate of the core radius in this
letter (eq.(\ref{eq:scaling})) is novel.

Here is an outline of the letter.  In the following section we review
the basic results on the flow of energy through the cusp around the
black hole.  Then we apply H\'enon's argument, which establishes the
way in which the radius of the core varies with the mass of the black
hole.  Next we compare our prediction with existing data from
simulations, and finally discuss the place of our result within the
literature on this subject.

\section{The Energy Flux from a Central Black Hole}

We consider a cluster of mass $M$ containing a central black hole of
mass $\mbh$.  The black hole is surrounded by a cusp, which merges at
its edge into a core of nearly constant density (Fig.\ref{fig:fit}).
Beyond the core radius the density falls off again in the halo of the
star cluster.

Inside the cusp,  the flux
of energy at radius $r$ is 
\[
\evar \sim \frac{\rho r^3 v^2}{\tr}.
\]
where $\rho$ is the stellar density, $v^2$ is the mean square stellar
velocity, and $\tr$ is the relaxation time.   This is given by 
$\tr\sim
\displaystyle{\frac{v^3}{G^2m\rho\ln\Lambda}}$, where $\ln\Lambda$ is
the Coulomb logarithm and $m$ is the individual stellar mass.  Hence
\[
\evar \sim \frac{G^2m\rho^2r^3\ln\Lambda}{v}.
\]
 In the part of the cluster where the potential is dominated
by the black hole, i.e. $r < \displaystyle{\frac{G\mbh}{v_c^2}}$,
where $v_c$ is the velocity dispersion in the core of the cluster, we
have $v^2\sim\displaystyle{\frac{G\mbh}{r}}$, and so
\[
\evar \sim \frac{G^{3/2}m\rho^2r^{7/2}\ln\Lambda}{\mbh^{1/2}}.
\]
If the flux is independent of $r$ \citep{LS1977} we get $\rho\propto r^{-7/4}$.

At the edge of the cusp we have $v^2 \sim v_c^2$, and so the radius of
the cusp is
\begin{equation}\label{eq:rcusp}
\rcusp\sim\displaystyle{\frac{G\mbh}{\vc^2}}.
\end{equation}
At this radius we also have $\rho\sim\rho_c$, where $\rhoc$ is the
density in the core, and so
\begin{equation}
\evar \sim \frac{G^5m\rhoc^2\mbh^3\ln\Lambda}{\vc^7}.
\end{equation}

\section{The Radius of the Core}

In steady post-collapse expansion this must balance the energy flux at
the half-mass radius, which is
\begin{eqnarray}
\evarh&\sim& \frac{M\vh^2}{\trh}\\
&\sim& \frac{G^2mM\rhoh\ln\Lambda}{\vh},
\end{eqnarray}
where the subscript $h$ denotes conditions at the half-mass radius.
Estimating $\rhoh\sim \displaystyle{\frac{M}{\rh^3}}$ and equating
$\evarh$ to
$\evar$, we obtain 
\[
\frac{M^2}{\vh\rh^3} \sim \frac{G^3\rhoc^2\mbh^3}{\vc^7}.
\]
If conditions are approximately isothermal between the core radius,
$r_c$, and the
half-mass radius, we can estimate
$\rhoc\sim\displaystyle{\frac{M}{\rh^3}\frac{\rh^2}{\rc^2}}$ and
$\vc^2 \sim \vh^2 \sim \displaystyle{\frac{GM}{\rh}}$, whence
\begin{equation}\label{eq:scaling}
\frac{\rc}{\rh} \sim \left(\frac{\mbh}{M}\right)^{3/4}.
\end{equation}

This scaling of core radius with black hole mass is expected to be
approximately valid provided that the resulting core radius exceeds
that of the cusp around the black hole (eq.\ref{eq:rcusp}).
Using the above estimate for $\vc^2$, it follows that 
\begin{equation}
\frac{\rcusp}{\rc}\sim \left(\frac{\mbh}{M}\right)^{1/4}.\label{eq:rcusp_on_rc}
\end{equation}
For a sufficiently massive black hole, the
density profile may show no sign of a core radius.  This conclusion
was already reached by \citet{MS1979}, though they
modelled systems with a core radius independent of the black hole
mass.  {A fit to numerical data (Sec.\ref{sec:fit}), however, 
yields a coefficient of about $0.7$ in eq.(\ref{eq:rcusp_on_rc}), and 
we find that, even up to a black hole mass of order 10\% of the
cluster mass, the radius of the cusp is less than half the core
radius.}  On the other hand unless $\mbh$ is large enough, the cusp contains too few
stars to provide a signature of the presence of the black hole \citep{Ba2005}.

\section{Comparison with Simulations}\label{sec:fit}

\begin{figure}
\caption{Example of a fit of eq.(\ref{eq:formula}) to the density
profile.  Data outside $r=2$ were not included in the fit, which is
for the case $\mbh/M = 0.0075$.  The dotted line shows the profile
without the contribution from the cusp, and in this figure the ``cusp radius'' is defined to be the radius at which the contributions from the core and cusp are equal.}
\includegraphics[angle=-90,width=9cm]{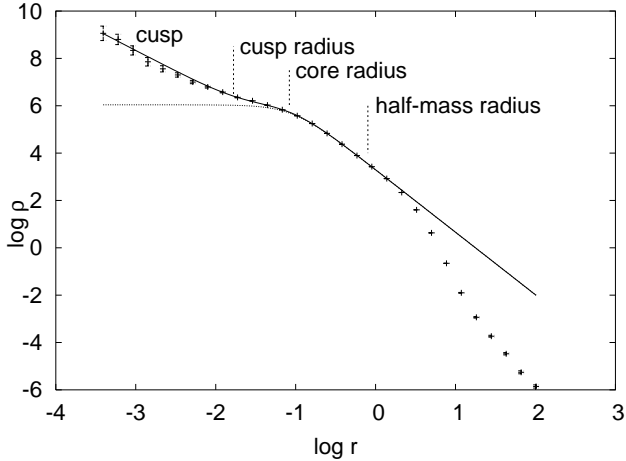}
\label{fig:fit}
\end{figure}

\citet{Ba2004a} have carried out a series of $N$-body simulations with
stars of equal mass and several different values for the initial black
hole mass.  In addition, unpublished data from some of these runs
allow us to measure the core radius for intermediate values of the
black hole mass, which increases during the course of each simulation.
The measurement of core radius is not entirely straightforward,
however.  While the $N$-body code itself gives a current value of the
``core radius'' \citep{Aa2003}, this is based on a density-weighted
average of the distances of stars from the density centre.  Because
the computed density around the black hole increases with increasing
black hole mass, this approach introduces a mass-dependent bias which
leads to an underestimate of the true core radius for large black hole
masses.  \citet{Ba2005} adopted a more observational procedure, and estimated
the core radius as the radius at
which the surface density drops to half its central value; but this is also
biased in the same way.
Therefore we have taken a different approach, in which we
fit the density distribution in the simulation by a template which
describes the cusp around the black hole embedded in a core.
Specifically, we fitted the density at radius $r$ by 
\begin{equation}\label{eq:formula}
\rho(r) =
(ar^{-7/4} +b)(1+r^2/r_c^2)^d,
\end{equation}
where $a,b,r_c,d$ are parameters.  For
$a = 0$ (no cusp) this resembles closely the so-called EFF model for
the distribution of density in a star cluster with a core
\citep{EFF1987,MG2003}.  The fit is not good at large radii; by
experimenting, we found that the mean square residual was minimised if
data outside about $2r_h$ were rejected (Fig.\ref{fig:fit}).

\begin{figure}
\caption{Dependence of core radius (in units of the half-mass radius)
on black hole mass (in units of the total cluster mass).  The line is
the graph of $\rc/\rh = 4.3(\mbh/M)^{3/4}$ (cf.
eq.\ref{eq:scaling}), which is the best fit.  Boxes: data from the four profiles in \citet{Ba2004a};
crosses: data at various times during the run with $\mbh(0) =
266M_\odot$ (though the initial data point is not plotted).  The error bars are the standard error provided by the
fitting package, and include the contribution  of sampling errors in the
individual density measurements.
}
\includegraphics[angle=-90,width=9cm]{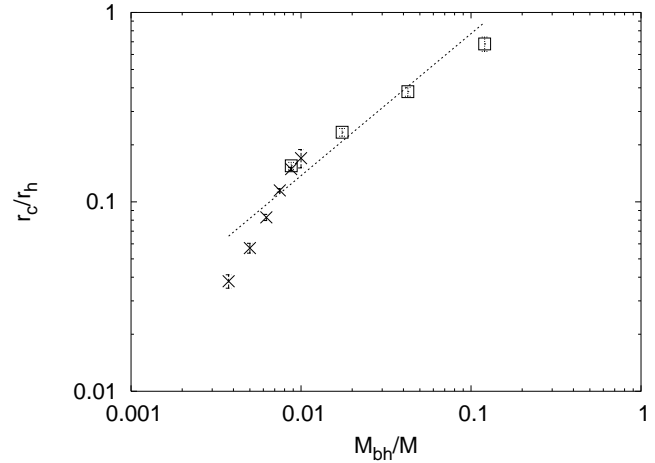}
\label{fig:radii}
\end{figure}

Figure \ref{fig:radii} shows fits of our theory to the data we have
obtained in this way.  {The points on the left (crosses) come from one
simulation, and plot the evolution of $r_c/r_h$ as the black hole
grows.  The early points in particular do not fit the predicted power
law very well, and we consider that these correspond to the period
during which equilibrium has not yet been established between the flux
of energy at the half-mass radius and that provided by the cusp.
The points on the right (boxes) represent four different
simulations, with different initial black hole mass, late in the
evolution.  The agreement with our theory is better, but the theory
depends implicitly on a homology assumption, and the slightly
discrepant slope 
may be due to small departures from homology;  this is not unexpected,
given that the values of $r_c/r_h$ extend up to about $0.8$.}

The important result to be obtained from this figure is that the
core radius {\sl increases} with black hole mass, which is the opposite of the
conclusion that would be drawn from Table 1 of
\citet{Ba2005}\footnote{We also note that the last value of
$R_C/R_{h,{\rm pro}}$ in that Table should be $0.07$.}.  There the core
radius was estimated from the radius at which the surface density
drops to half its central value (see above), and the fact that their
paper considered clusters with a stellar mass function is not relevant
in the present context.  \citet{TAMH2007} found that the dependence of
core radius on $\mbh$ is flatter than we predict, but there the quoted
core radius was estimated as in $N$-body models (see above).

\section{Discussion}\label{sec:discussion}

In this Letter we have argued that  a balance is reached between the
energy production of the cusp and the energy required for expansion of
the entire cluster.  Previously \citet{MS1980} considered
that there should be a balance between energy production of the cusp
and that required for expansion of the {\sl core}.  This, however,
leads to the relation $\rc/\rcusp \sim$ constant, independent of the
mass of the black hole and cluster.  This is certainly incorrect when
the the black hole mass is very small.  \citet{Sh1977}, \citet{McM1981} and \citet{DS1982} 
also, in effect, equated the generation of energy in the
cusp to the energy required to expand the core; in fact, however, most
of the energy passes through the core to the half-mass radius, as it
provides the energy needed for the expansion of the entire cluster.
\citet{Ba2004a} equated the energy generation to that required at
$r_h$, but assumed a constant value of $r_h/r_c$.
\citet{YZ1990} and \citet{Do1991} took a homology model for
the stellar system, which was thus characterised by a single scale
radius.

The theory we have presented depends on a number of assumptions.
\begin{enumerate}
  \item It applies only to systems with stars of
equal mass, {though there is no reason to suppose that it does not
extend, with suitable modification of detail, to systems with a
realistic mass spectrum}. 
\item  It also depends, though in an inessential
way, on the assumption that $v_c^2\sim v_h^2$.  In fact $N$-body
models, albeit with a mass spectrum \citep{Ba2005}, show that the
$v^2$ varies by only about 30\% between $r_h$ and $0.1r_h$.  If
instead we had assumed that $v^2\propto r^{-\beta}$, where these
results imply that $\beta\sim 0.1$, our result (eq.(\ref{eq:scaling}))
would have changed from a power law index of $3/4$ to $6/(8 -
3\beta)\simeq 0.78$.  The velocity dispersion profiles of some
clusters require larger values of $\beta\sim0.3$, however (e.g. $\omega$ Cen
\citep{MMDD1995}, M15 \citep{MHA2003}).
\item The theory         applies only when sufficient time has elapsed for achieving a balance
between the energy produced in the cusp and the flow of energy across
the half-mass radius.  In almost all galactic nuclei, and even in many
globular clusters \citep{T2006}, the relaxation time
would  be too long for the conditions assumed in this Letter to be
established. 
\item The theory does not apply to systems where some other mechanism,
  e.g. interactions of primordial  binaries, provide energy more efficiently.
\item It does not apply to systems that are so dense
(including  many
galactic nuclei) that the structure of the cusp is dominated by the
role of physical collisions, and the power law is altered
\citep{Ra1999,DS1983}. 
\end{enumerate}

\section*{Acknowledgements}

DCH, PH and JM thank SM for his hospitality during a visit to the
Yukawa Institute, Kyoto, where much of the work was carried out.  This
work was supported in part by the Grants-in-Aid of the Ministry of
Education, Science, Culture, and Sport (14079205, 16340057 S.M.), and
by the Grant-in-Aid for the 21st Century COE ``Center for Diversity
and Universality in Physics'' from the Ministry of Education, Culture,
Sports, Science and Technology (MEXT) of Japan.  We are grateful to
M. Trenti and the referee, F.A. Rasio, for their comments.


\begin{thebibliography}{}

\bibitem[Aarseth(2003)]{Aa2003} Aarseth, S.~J.\ 2003, 
Gravitational N-Body Simulations, 
Cambridge, UK: Cambridge University Press

\bibitem[Amaro-Seoane et al.(2004)]{AS2004} Amaro-Seoane, P., 
Freitag, M., \& Spurzem, R.\ 2004, \mnras, 352, 655 

\bibitem[Arabadjis(1997)]{Ar1997} Arabadjis, J.~S.\ 1997, 
Ph.D.~Thesis,  University of Michigan

\bibitem[Bahcall \& Wolf(1976)]{BW1976} Bahcall, J.~N., \& 
Wolf, R.~A.\ 1976, \apj, 209, 214 

\bibitem[Baumgardt et al.(2005)]{Ba2005} Baumgardt, H., 
Makino, J., \& Hut, P.\ 2005, \apj, 620, 238 

\bibitem[Baumgardt et al.(2004a)]{Ba2004a} Baumgardt, H., 
Makino, J., \& Ebisuzaki, T.\ 2004a, \apj, 613, 1133 

\bibitem[Baumgardt et al.(2004b)]{Ba2004b} Baumgardt, H., 
Makino, J., \& Ebisuzaki, T.\ 2004b, \apj, 613, 1143 


\bibitem[Cohn \& Kulsrud(1978)]{CK1978} Cohn, H., \& Kulsrud, 
R.~M.\ 1978, \apj, 226, 1087 

\bibitem[Dokuchaev(1991)]{Do1991} Dokuchaev, V.~I.\ 1991, 
\mnras, 251, 564 


\bibitem[Duncan \& Shapiro(1982)]{DS1982} Duncan, M.~J., \& 
Shapiro, S.~L.\ 1982, \apj, 253, 921 

\bibitem[Duncan \& Shapiro(1983)]{DS1983} Duncan, M.~J., \& 
Shapiro, S.~L.\ 1983, \apj, 268, 565 

\bibitem[Elson et al.(1987)]{EFF1987} Elson, R.~A.~W., Fall, 
S.~M., \& Freeman, K.~C.\ 1987, \apj, 323, 54


\bibitem[Freitag \& Benz(2002)]{FB2002} Freitag, M., \& Benz, 
W.\ 2002, \aap, 394, 345 

\bibitem[H{\'e}non(1975)]{He1975} H{\'e}non, M.\ 1975, IAU 
Symp.~ 69: Dynamics of Stellar Systems, 69, 133


\bibitem[Lightman \& Shapiro(1977)]{LS1977} Lightman, A.~P., 
\& Shapiro, S.~L.\ 1977, \apj, 211, 244

\bibitem[Mackey \& Gilmore(2003)]{MG2003} Mackey, A.~D., \& 
Gilmore, G.~F.\ 2003, \mnras, 338, 85

\bibitem[McMillan et al.(1981)]{McM1981} McMillan, S.~L.~W., 
Lightman, A.~P., \& Cohn, H.\ 1981, \apj, 251, 436


\bibitem[McNamara et al.(2003)]{MHA2003} McNamara, B.~J., 
Harrison, T.~E., \& Anderson, J.\ 2003, \apj, 595, 187

\bibitem[Marchant \& Shapiro(1979)]{MS1979} Marchant, A.~B., 
\& Shapiro, S.~L.\ 1979, \apj, 234, 317 

\bibitem[Marchant \& Shapiro(1980)]{MS1980} Marchant, A.~B., 
\& Shapiro, S.~L.\ 1980, \apj, 239, 685 

\bibitem[Meylan et al.(1995)]{MMDD1995} Meylan, G., Mayor,
  M., 
Duquennoy, A., \& Dubath, P.\ 1995, \aap, 303, 761

\bibitem[Murphy et al.(1991)]{Mu1991} Murphy, B.~W., Cohn, 
H.~N., \& Durisen, R.~H.\ 1991, \apj, 370, 60 

\bibitem[Rauch(1999)]{Ra1999} Rauch, K.~P.\ 1999, \apj, 514, 
725 

\bibitem[Shapiro(1977)]{Sh1977} Shapiro, S.~L.\ 1977, \apj, 
217, 281 

\bibitem[Shapiro \& Lightman(1976)]{SL1976} Shapiro, S.~L., \& 
Lightman, A.~P.\ 1976, \nat, 262, 743 

\bibitem[Shapiro \& Marchant(1978)]{SM1978} Shapiro, S.~L., \& 
Marchant, A.~B.\ 1978, \apj, 225, 603 

\bibitem[Trenti(2006)]{T2006} Trenti, M.\ 2006, ArXiv 
Astrophysics e-prints, arXiv:astro-ph/0612040 

\bibitem[Trenti et al.(2007)]{TAMH2007} Trenti, M., Ardi,
  E., 
Mineshige, S., \& Hut, P.\ 2007, \mnras, 374, 857

\bibitem[Yuan \& Zhong(1990)]{YZ1990} Yuan, Z., \& Zhong, 
X.~G.\ 1990, \apss, 168, 233 


\end{thebibliography}
\end{document}